# Ultralow voltage, High-speed, and Energy-efficient Cryogenic Electro-Optic Modulator

Paolo Pintus,[1,2,‡,*] Anshuman Singh,[3,‡] Weiqiang Xie,[1,‡] Leonardo Ranzani,[3] Martin V. Gustafsson,[3] Minh A. Tran,[1,4] Chao Xiang,[1] Jonathan Peters,[1] John E. Bowers,[1] Moe Soltani[3,*]

Affiliations:
[1]Department of Electrical and Computer Engineering, University of California Santa Barbara, California 93106, USA
[2]Department of Physics, University of Cagliari, Monserrato, 09042, Italy
[3]Raytheon BBN Technologies, 10 Moulton Street, Cambridge, Massachusetts 02138, USA
[4]Currently with Nexus Photonics, Goleta, California 93117, USA

[*]Corresponding authors: ppintus@ece.ucsb.edu, mo.soltani@raytheon.com

[‡]These authors contributed equally to the work

**Abstract**:

Photonic integrated circuits (PICs) at cryogenic temperatures enable a wide range of applications in scalable classical and quantum systems for computing and sensing. A promising application of cryogenic PICs is to provide optical interconnects by up-converting signals from electrical to optical domain, allowing massive data-transfer from 4 K superconducting (SC) electronics to room temperature environment. Such a solution is central to overcome the major bottleneck in the scalability of cryogenic systems, which currently rely on bulky copper cables that suffer from limited bandwidth, large heat load, and do not show any scalability path. A key element for realizing a cryogenic-to-room temperature optical interconnect is a high-speed electro-optic (EO) modulator operating at 4 K with operation voltage at mV scale, compatible with SC electronics. Although several cryogenic EO modulators have been demonstrated, their driving voltages are significantly large (several hundred mV to few V) compared to the mV scale voltage required for SC circuits. Here, we demonstrate a cryogenic modulator with ~10 mV peak-to-peak driving voltage and gigabits/sec data rate, with ultra-low electric and optical energy consumptions of ~10.4 atto-joules/bit and ~213 femto-joules/bit, respectively. We achieve this record performance by designing a compact optical ring resonator modulator in a heterogeneous InP-on-Silicon platform, where we optimize a multi-quantum well layer of InAlGaAs to achieve a strong EO effect at 4 K. Unlike other semiconductors such as silicon, our platform benefits from the high-carrier mobility and minimal free carrier freezing of III-V compounds at low temperatures, with moderate doping level and low loss (intrinsic resonator Q~272,000). These modulators can pave the path for complex cryogenic photonic functionalities and massive data transmission between cryogenic and room-temperature electronics

1. **Introduction**

A class of next generation classical and quantum computing systems[1–4] and sensors[5,6] will operate at cryogenic temperatures (~4 K) based on superconducting and optical technologies. A major challenge in these platforms is providing high-speed and energy efficient transfer of massive amounts of data (e.g., hundreds of gigabits/s to terabits/sec) to room temperature. Commonly used coaxial cables have limited communication bandwidth, large propagation loss, and add excessive heat load to the cryogenic stage[3,7]. Thus, they will not be a scalable solution for the data transmission from cold to room temperature, as the heat load counteracts the energy efficiency gained by using superconducting and other low-temperature circuits, and can easily exceed the cooling power of available cryocoolers as systems scale. To overcome these limitations, a promising solution is to use integrated electro-optic (EO) modulators in a photonic integrated circuit (PIC) platform to up-convert and multiplex data from the electrical to the optical domain and transmit them to room temperature via optical fibers[3,7,8]. This approach provides wide data bandwidth with a potential aggregate data capacity of tens of terabits/s to petabits/s [3,8,15], extremely low loss, and negligible heat transfer. To achieve this goal, the electro-optic (EO) modulator needs to operate at a temperature of ~4 K and responds to high- speed, low-voltage SC digital signals, typically on the scale of millivolts[5,10–12].

There have been several recent demonstrations of cryogenic EO modulators based on different material platforms [7,9–14]. Despite this progress, however, the demonstrated EO platforms operate at modulation voltages 2-3 orders of magnitude higher than the typical sub-10 mV voltages provided by SC circuits. In addition, not all EO platforms can operate reliably at low temperatures. Common EO materials such as polymers and Lithium Niobate are prone to photorefractive instability at low temperatures, making them unreliable for interfacing with a SC circuitry[18,19]. Long-established silicon EO modulators based on free-carrier plasma dispersion become inefficient below a few tens of kelvin due to free-carrier freeze-out[20]. Although this effect can be compensated to some degree by increasing the doping concentration in the silicon, it does not improve the voltage sensitivity of the modulator[7,17] nor does it address the increased insertion loss due to excess free carriers. A recent promising cryogenic Si modulator paper[21] adds a cryogenic voltage amplifier stage before the modulation stage to provide the large modulation voltage required for the modulator stage, but at the cost of substantially increasing the total energy consumption by the cryogenic amplifier. Silicon modulators based on DC-Kerr induced Pockels effect have been demonstrated at 4 K, however the required modulation signal is on the scale of volts[11]. Plasmonic slotted waveguide modulators infiltrated with EO polymer[22,23] have been demonstrated at room temperature with high-speed and small electric energy consumption at sub-100 aJ/bit [22,23], but they operate at several hundred millivolt modulation and suffer from large optical material loss, as well as large coupling loss due to mode mismatch between the straight ridge waveguide region and the slotted waveguide. Additionally, the material stability of EO polymers at 4 K temperature is not well known and susceptible to degradation over time. 2D material-based modulators[14] have been investigated at 4 K, but the modulation voltage is still more than two orders of magnitude higher than what is required to interface with SC circuitry.

Here, we demonstrate a Cryogenic InP-on-Si Photonic (CRISP) ring modulator (Fig. 1) which achieves ~10 mV peak-to-peak modulation voltage (10 mV$_{pp}$) at gigabits-per-second (Gbits/s) data rate, with electrical and optical energy consumption of 10.4 atto-joule-per-bit (aJ/bit) and 213 femto-joules-per-bit (fJ/bit), respectively, at 4 K temperature. The CRISP modulator is fabricated in a heterogeneous platform which combines the strong EO properties of InP – and its associated III-V alloys – with scalable and compact silicon photonic circuitry[24–27]. Unlike silicon modulators, the CRISP modulator suffers minimally from free-carrier freezing at low temperatures, allowing it to operate efficiently with moderate doping levels in the InP layer, thus enabling high Q resonators. We design the modulators for operation at 4K and a wavelength range within 1500-1600 nm.

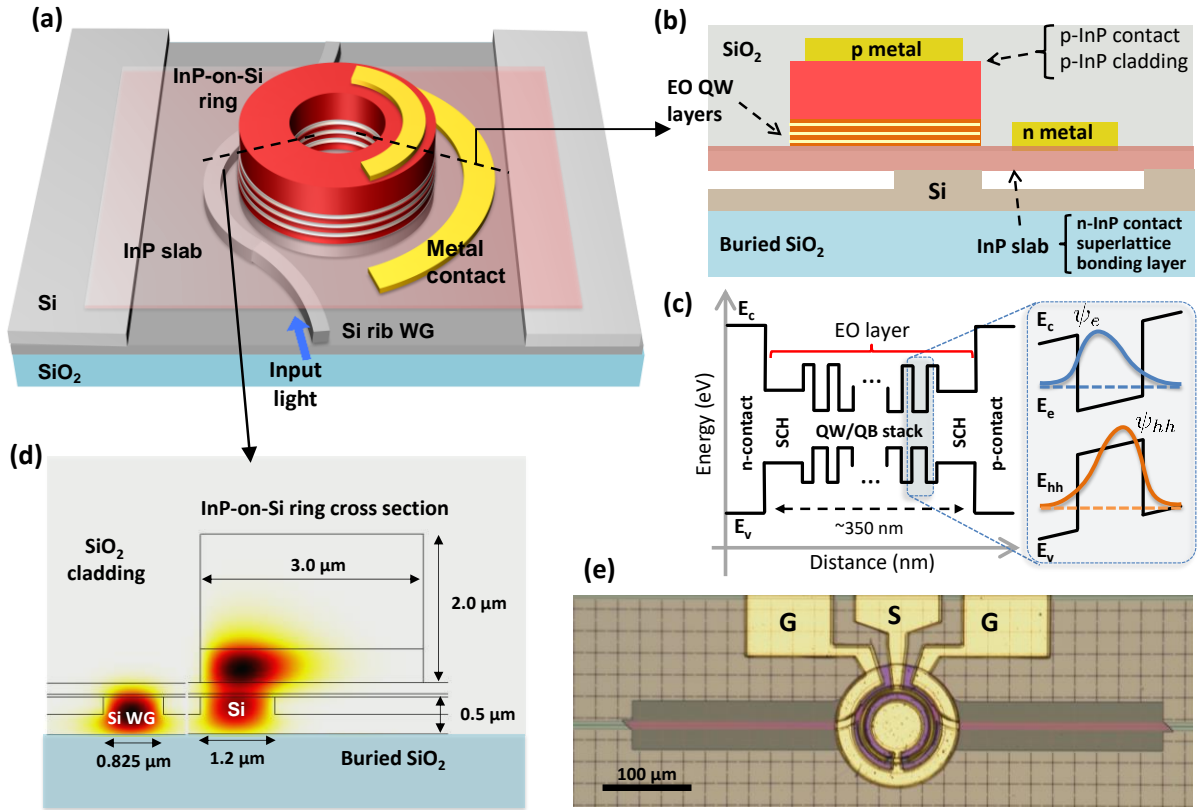

**Figure. 1 Heterogeneously integrated InP-on-Si resonator modulator for operation at 4 K**. (a) InP ring incorporating a stack of quantum wells bonded to a silicon ring, forming an InP-on-Si ring resonator which is coupled to a silicon waveguide. (b) Cross section of the stack of materials for the InP-on-Si modulator with metal contacts. (c) Energy band and size of the electro-optic (EO) layer which contains stacks of quantum wells (QWs) and quantum barriers (QBs) in an alternating pattern, buried between two separate confinement heterostructure (SCHs). The inset shows the wave functions of electrons ($\psi_e$) and heavy holes ($\psi_{hh}$) in the conduction and valence band, respectively. The conduction energy level ($E_c$) and the valence energy level $E_v$ are shown with respect to distance from the Silicon waveguide interface. Doping of the layers is optimized to achieve large band-filling and plasma effects, while QW geometry is engineered to maximize electron-hole interaction (exciton) to produce a strong quantum confined Stark-effect (QCSE). (d) Cross-section of optical mode profiles for the InP-on-Si resonator with a nominal external radius of 42 µm, and for the Silicon waveguide, showing the mode overlap between the waveguide and resonator modulator. (e) Optical microscope image of the fabricated ring modulator, showing metallic pads in ground-signal-ground (GSG) configuration for electric signals and silicon waveguide in pulley coupling arrangement with respect to the ring resonator.

## 2. Device Design and Characterization

A schematic of the CRISP modulator is shown in Fig. 1a where the stack of III-V compounds sits on a silicon ring which is evanescently coupled to a silicon rib waveguide. Two strong modulation mechanisms are present in the III-V alloys used in the modulator: band-filling[28] (Burstein-Moss) and quantum confined Stark effects[29] (QCSE), with some contributions also from plasma dispersion[28]. We design a multi-quantum well (QW) layer to enhance the EO effects and leading to large modulation. To further increase the modulation sensitivity to millivolt-scale voltages, we take advantage of a ring resonator configuration with a sharp resonance spectrum. This enables intensity modulation of the light with modulation voltages as small as ~10 mV$_{pp}$ and even lower.

A key aspect of the CRISP modulator is the design of the multi QW layer to achieve high-sensitivity modulation at 4 K. The doping of the QW layer is optimized to achieve large band-filling and plasma effects with negligible optical absorption, while the III-V alloy composition and the geometry of the QWs are engineered to maximize the QCSE. Specifically, the QCSE is the interaction of light with exciton quasiparticles (electron-hole bound states) and it becomes significantly stronger at lower temperatures due to the sharper exciton spectrum[29].

A simplified InP-on-Si cross-section of the modulator is shown in Fig. 1(b), where the EO layer consists of a stack of fifteen alternate QWs and quantum barriers (QBs). Each QW has a thickness of 8 nm sandwiched between 5 nm thick QB. This multi-QW/QB layer is also inserted between two 125 nm thick separate confinement heterostructures (SCHs) that confine electrons and holes in the EO layer. Such a design allows for a large overlap between electron and hole wave functions (Inset in Fig. 1(c)), so the QCSE effect is maximized. The layers of QW, QB and SCH are made of In$_x$Al$_{1-x-y}$Ga$_y$As, which provides a larger bandgap offset – useful for energy band engineering[30] – and higher fabrication yield than In$_{1-x}$Ga$_x$As$_{1-y}$P$_y$.

We optimize the dimensions of the silicon and the InP layer for maximum optical modal overlap (Fig. 1d). For this purpose, we choose a silicon layer with a thickness of 500 nm, and etch it down to 260 nm to produce a rib type waveguide for maximum optical index matching to the InP layer. For the modulator device fabrication (see Fig. 1e), we design a ring with an external radius of 42 µm and a Si ring width of 1.0 µm and 1.2 µm (see section S1 of the Supplementary material for the detailed design discussion).

Figure 2 shows the details of the III-V epitaxial layer that we designed for the CRISP modulator. The quaternary composition of QW, QB and SCH is chosen to maximize the EO effects in our modulator for 4 K operation according to the following considerations: (i) the lattice constants match InP to avoid undesirable strain when the device is cooled down to 4 K, (ii) the exciton energy (i.e., photoluminescence) in QWs is ~100 meV larger than the energy of the operating wavelength (λ~1500 - 1600 nm) , i.e., ~800 meV to prevent optical absorption[31], (iii) the energy gap and bandgap offset of QW and QB are designed to better confine electrons and holes in the QW to maximize QCSE as well as the band-filling and free-carrier dispersion effect[32,33]. By considering all these aspects, the alloy composition $In_xAl_{1-x-y}Ga_yAs$ used for QW, QB and SCH are chosen to be (x = 0.5296, y = 0.4512), (x = 0.5296, y = 0.3130) and (x = 0.5296, y = 0.3365), respectively (see section S2 the Supplementary material for more details on the material stack optimization). In addition, the three regions are lightly n-doped with a concentration of ~$1·10^{17}$/cm$^3$ in order to obtain a larger change in the refractive index of $In_{1-x-y}Al_xGa_yAs$ caused by plasma and band-filling effects[34].

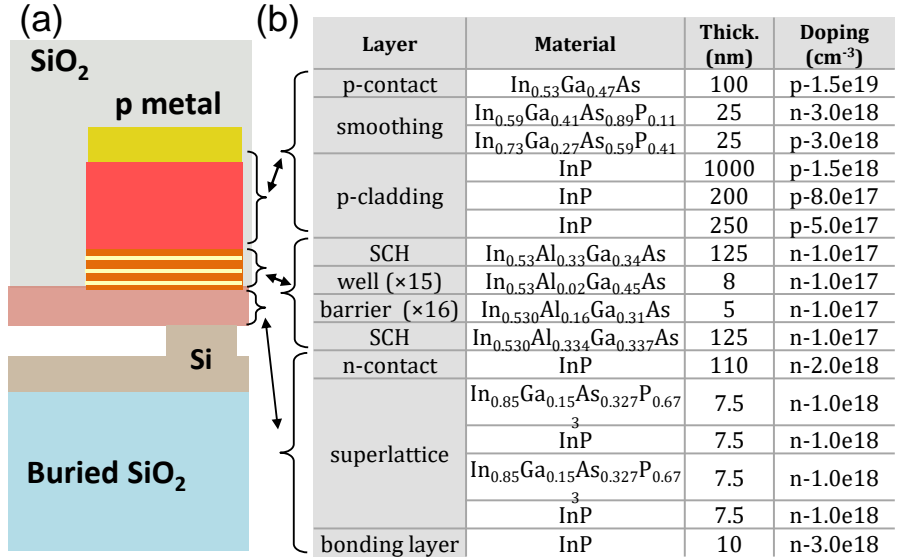

**Fig. 2**: **III-V epitaxial layer designed and optimized for the 4 K operation of the CRISP modulator**. (a) Schematic cross section of the CRISP ring resonator. (b) Epitaxial layer and their parameters.

A microscope image of the final device is shown in Fig. 1e, where the top view of the gold electrodes and the rib waveguide bus are clearly identifiable. A 2 μm thick SiO₂ cladding is under the Si layer, and a 1.5 μm thick SiO2 is the top cladding of the device (see Supplementary Section S3 for the details of the device fabrication process).

We characterize the EO performance of the CRISP modulator at 4 K in a cryogenic probe station. The fabricated CRISP modulator devices show a fiber-to-chip coupling loss of ~12 dB/facet, which can be largely reduced to 2.5 dB/facet by optimizing the fabrication process[35]. At the output of the chip, we use optical and electronic amplifiers at room temperature to amplify the received signal for the measurement (see sections S4-S6 of the Supplementary material for more details on the measurement).

Figure 3a shows the optical spectrum of the modulator, around 1600 nm, when the voltage sweeps between 0 V and -3.0 V with a step of 1 V. The resonator modulator has a loaded quality factor of ~53,000 corresponding to a resonance linewidth of ~30 pm (~3.75 GHz). With this small resonance linewidth, a minimal applied voltage is sufficient to bring the ring out of resonance. This large sensitivity to an electrical signal is shown in Fig. 3b, where the resonance shift is plotted with respect to the applied reverse bias voltage. When the magnitude of the reverse bias is small, the spectrum shifts mainly because of the band-filling and the carrier depletion effects ($\Delta\lambda \propto V^{1/2}$). For much lower bias voltages, the QCSE becomes the dominant contribution ($\Delta\lambda \propto V^{3/2}$), explaining the non-uniform shift of the spectrum as a function of the applied voltage[34]. From Fig. 3b we choose the optimal voltage bias point to provide large modulation sensitivity with optimum optical performance. Though the resonance shift is much stronger in the QCSE regime, we biased the modulator such that the band-filling and plasma effects are dominant since we observed negligible absorption and no resonance linewidth broadening in this region. An optimal future design can potentially bring the QCSE to a region of lower absorption while providing large resonance shift.

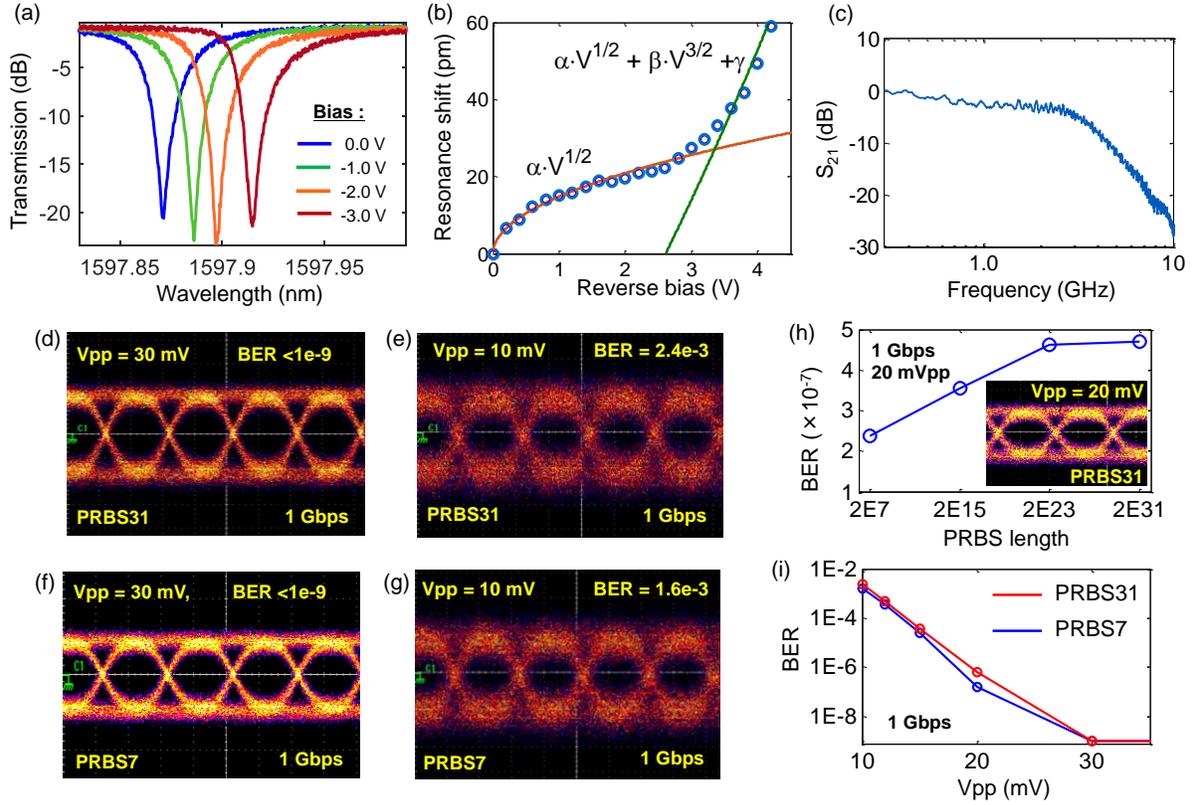

**Fig. 3**: **CRISP modulator optimized for 10 mV scale operation at 4 K**. (a) Spectrum of the resonator modulator at different reverse voltage biases, with a full-width at half maximum (FWHM) for the resonance linewidth of ~30 pm. (b) Wavelength shift of the modulator resonance with respect to reverse bias voltage. The fits (lines) of measured data (circle) highlight the contributing modulation mechanisms. The shift is proportional to $V^{1/2}$ for the band-filling and the plasma effects, while it varies with $V^{3/2}$ for the QCSE. The constants $\alpha = 14.79 \text{ pm}/\sqrt{V}$, $\beta = 12.45 \text{ pm}/\sqrt{V^3}$, and $\gamma = -76.36$ pm are the fitting parameters. (c) The electro-optical frequency response ($S_{21}$ coefficient) of the modulator, shows a 3 dB bandwidth of ~4 GHz. (d)-(g) Demonstration of millivolt scale and 1 Gbps data rate modulation by sending PRBS $2^{31}-1$ and PRBS $2^7-1$ data stream to the modulator. For 30 mV, we achieve BER<1e-9 limited by the BER floor of our equipment. The insets in (d)-(g) show the applied peak-to-peak modulation voltage ($V_{pp}$), bit error rate (BER), PRBS length, and the data rate. In (d) and (f) the electric energy consumption is 92 aJ/bit, and in (e) and (g) it is 10.4 aJ/bit. The modulator optical energy consumption in (d)-(g) is 213 fJ/bit. (h) BER versus different PRBS lengths, for 1 Gbps and 20 mV modulation. (i) The BER of the input signal decreases by increasing the modulation voltage and the transmission becomes error free at 30 mV and above. The modulator bias voltage for these experiments is -2 V and the on-chip optical power sent to the modulator is 213 μW.

Our experiments show gigabits-per-second modulation rate with sub-10 mV$_{pp}$ modulation signal and ~10.4 aJ/bit electric energy consumption. The EO frequency response is shown in Fig. 3c and indicates a modulation bandwidth of around 4 GHz. This value is limited by the optical resonance bandwidth of ~3.5 GHz due to the large loaded quality factor of the ring resonator (Q ~53,000). For the modulator data rate measurement, we apply to the electrodes a pseudo-random non-return-to-zero (NRZ) bit sequence of length of $2^{31}$-1 and $2^{7}$-1, and peak-to-peak voltage amplitude (V$_{pp}$) at the millivolt scale. Figure 3d-3g show the measured eye diagrams for data rates of 1 Gbps and different modulation amplitudes ranging down to 10 mV$_{pp}$. The total electric energy consumption per bit is equal to $\frac{CV_{pp}^2}{4} + |V_{bias}I_{bias}|/r_b$, where C is the modulator capacitance, V$_{bias}$ is the bias voltage, I$_{bias}$ is the leakage current of the modulator, and r$_b$ is the data rate. We measure a capacitance of C ~ 406 fF at different bias voltages (see section S6 of the Supplementary), which results in $\frac{CV_{pp}^2}{4} \sim 10.1\ aJ/bit$, when V$_{pp}$ = 10 mV. For this modulator device we observe a static leakage current of 0.14 nA, resulting in an energy consumption of 0.3 aJ/bit for a bias of - 2 V and a data rate of 1 Gbps (see section S7 of the Supplementary material). Therefore, we estimate a total electric energy per bit E$_b$ = 10.4 aJ/bit. We believe that sub-aJ/bit electric energy consumption can be reached by reducing the electrode size and introducing an intrinsic region between the p and n doped areas, reducing the diode capacitance as well as the energy-per-bit[36]. Furthermore, we measured the bit error rate (BER) of all the eye diagrams shown in Fig. 3(d)-(g), which is indicative of the modulator performance for different input parameters such as bit sequence length, data rate, and modulation amplitude. We measure the BER in the range of $10^{-3}$ for $V_{pp}$ of 10 mV, and $< 10^{-9}$ (which is the noise floor of the BER equipment) for $V_{pp}$ of 30 mV for both PRBS $2^{31}$- and $2^7$-1 bit sequences. Figure 3(h) shows the BER for various bit sequence lengths for 20 mV$_{pp}$ of modulation amplitude, demonstrating similar BER ($10^{-7}$). Similarly, a plot of BER vs. $V_{pp}$ (Fig. 3(i)) shows an expected behavior where the BER decreases with increasing amplitude and has error-free operation above 30 mV$_{pp}$ up to PRBS $2^{31}$-1. With an on-chip optical power of 213 µW sent to the modulator, the optical energy consumption is 213 fJ/bit for these measurements.

Figure 4 shows further experiments with the modulator device of Fig. 3, but for higher data rates up to 2 Gbps, and with 30 mV$_{pp}$ modulation, showing error-free operation at these data rates for PRBS $2^7$-1. The BER vs $V_{pp}$ curve clearly indicates the increase in the BER for higher data rates, see Fig. 4(c). Although this modulator has a bandwidth of 4 GHz, the limited bandwidth of our electronic amplifier (1.3 GHz) used after the room temperature photodetector did not allow us to measure efficient eye diagrams beyond 2 Gbps for low modulation voltages.

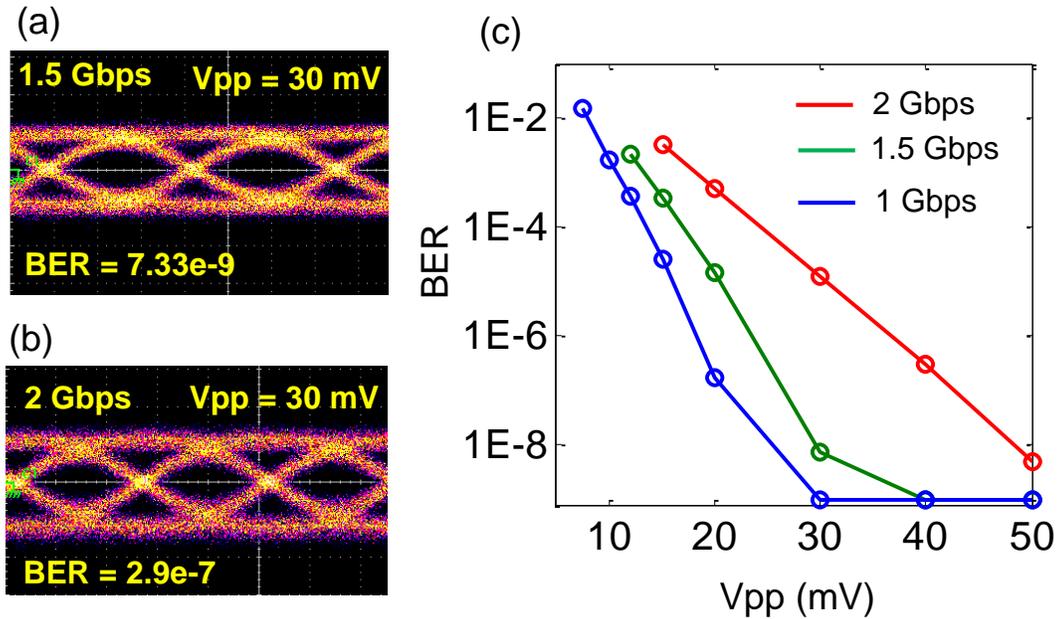

**Fig. 4: Bit error rates for data rates up to 2 GBPS at 4 K.** (a)Date rate of 1.5 and (b) 2 Gbps, respectively for PRBS $2^7$-1 for the modulator shown in Fig. 1. The electric energy consumption in both (a) and (b) is 92 aJ/bit, and the optical energy consumptions are 142 fJ/bit and 107 fJ/bit, respectively. (c) BER vs modulation voltages show increased error as the date rate goes up for a similar level of the modulation voltage. The bias voltage and on-chip power are 213 µW and -2 V, respectively.

One advantage of our CRISP modulator is its minimal free-carrier freezeout allowing wider bandwidth modulators at low temperatures. To verify that our modulators suffer minimally from carrier freezing and are able to operate at higher modulation data rates, we repeat the experiments with a lower-Q resonator that exhibits a larger resonance linewidth, and thereby a wide modulation bandwidth. For this experiment, we select a ring resonator with Q ~20,000 corresponding to a resonance linewidth of ~80 pm (Fig. 5a). Such a wider resonance linewidth requires a larger modulating voltage. Fig. 5(b) shows the EO frequency response with a 3 dB bandwidth of ~9 GHz, which is much wider than the results shown in Fig. 3(c). Note that we used a different electronic amplifier for this experiment (10 GHz bandwidth) after the photodetector. Figures 5(c)-5(d) show the measured eye diagrams at 4 Gbps data rates, for a modulation voltage of 110 mV$_{pp}$, and at different PRBS values ($2^7$- 1 and $2^{15}$-1). For this device, the electric energy-per-bit is ~1.3 fJ/bit which is still significantly small compared to the prior demonstrated works in the literature.

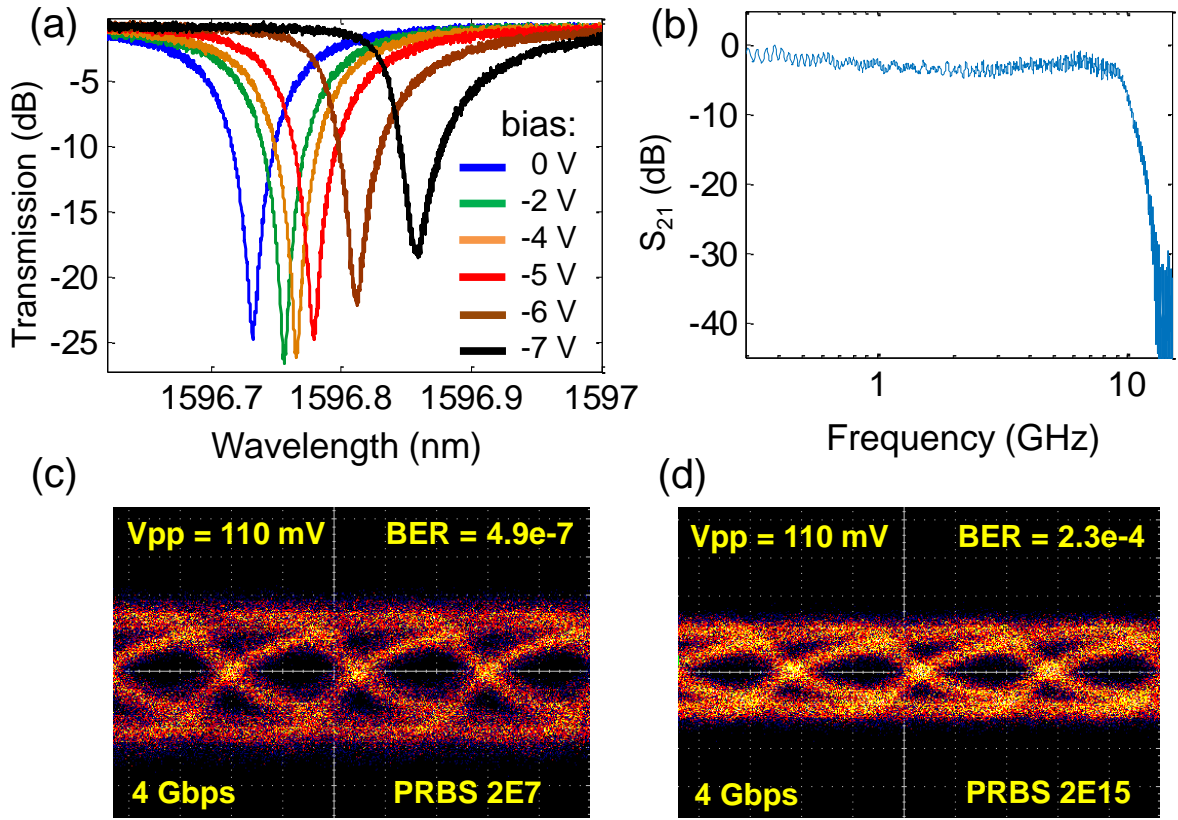

**Fig. 5: CRISP modulators with wider modulation bandwidth and higher data rates at 4 K.** (a) Spectrum of a resonator modulator with large resonance linewidth (~80 pm) and its shift at different bias voltages shown in the inset. (b) The measured modulator frequency response shows a 3 dB bandwidth of ~9 GHz. (c)-(d) The eye diagrams for data rates (modulation voltages) of 4 Gbps ($V_{pp}$ = 110 mV) for PRBS $2^7$-1 and PRBS $2^{15}$-1 respectively. The modulator is biased at $V_{bias}$ = -6 V ($I_{bias}$ = 0.08 nA), operating dominantly in the QCSE regime. The eye diagrams and bandwidth are obtained for on-chip optical power of 1.1 mW. The modulator electrical and the optical energy consumption is 1.3 fJ/bit and 275 fJ/bit, respectively.

Table 1 summarizes a comparison of the CRISP modulator with other demonstrated cryogenic EO modulators. As shown in the table, the CRISP modulator outperforms other EO modulators, operating at significantly smaller driving voltage and with minimal electrical energy consumption. In all the devices shown in Table 1, the optical energy consumptions of the modulators are almost in the same range and much larger than the electric energy consumption. Therefore, it is critical to reduce any optical dissipation and insertion loss in such modulators. Since dissipation is one of greatest detriment in the cryogenic environment, all other energy consuming elements including the lasers, and any optical and electrical amplifiers and detectors can be stationed at room temperature.

The radius of the CRISP resonator modulator in this work is 42 µm. In principle, there is room to reduce this radius which will increase the resonance shift sensitivity to smaller voltages as well as reduce the capacitance and increase the modulator bandwidth. However, proper designs of InP and Si layer dimensions are required to avoid large bending loss while shrinking the resonator radius. In this work, the highest intrinsic Q that we were able to achieve was ~272,000 (see section S8 of the Supplementary material). The large intrinsic Q of such resonator modulators can enable other applications including sensing at cryogenic temperature. This is an advantage enabled by high carrier mobility and minimal free-carrier freezing of the InP layer, resulting in moderate doping requirement and with lower loss, while conventional Si modulators have their Q limited to ~$10^4$ due to high-doping induced loss.

**Table 1**: A comparison of CRISP modulator with other demonstrated cryogenic EO modulators

| Material and modulator struct. | BW (GHz) | Bit Rate (Gbps) | Modulation voltage (Vpp) | Electrical Energy | Optical Energy | Ref. |
|---|---|---|---|---|---|---|
| LiNbO3 waveguide phase modulator | 5 | 5 | 5 V | - | - | 16 |
| BaTiO3 microring, Pockels effect | 30 | 20 | 1.7 V | 45 fJ/bit | - | 13 |
| Graphene on SiN microring | 14.7 | - | - | - | - | 14 |
| Si microdisk | 4.5 | 10 | 1.8 V | - | 80 fJ/bit | 7 |
| Si MZI modulator, DC Kerr effect | 1.5 | -- | - | - | 5 pJ/bit | 15 |
| Si ring modulator | 2.5 | 5 | 0.5 V | - | - | 17 |
| Si ring modulator | 2.5 | 20 | 1.5 V | - | - | 17 |
| Si ring modulator with electric pre-amp. | 2.2 | 1 | 100 mV (4 mV)** | 50 fJ/bit* | 320 fJ | 21 |
| Si ring modulator with electric pre-amp. | 2.2 | - | 250 mV (10 mV)** | 300 fJ/bit* | 800 fJ | 21 |
| **InP-on-Si ring QW modulator Device # 1 (Fig. 3)** | 4 | 1 | 10 mV | 10 aJ/bit | 210 fJ | **This work** |
| **InP-on-Si ring QW modulator Device #2 (Fig. 5)** | 9 | 4 | 110 mV | 1.3 fJ/bit | 275 fJ | **This work** |

\*: This is the consumed power by the cryogenic pre-amplifier before the modulator

\**: The numbers in the parenthesis are the voltage applied to the pre-amplifier with ~ 28 dB gain.

## 3. Conclusion

In summary, we have demonstrated a high-speed InP-on-Silicon electro-optic resonator modulator operating at 4 K with record performance of modulating voltage – as low as 10 mV$_{pp}$ – and electric energy consumption – as low as 10.4 aJ/bit. Those record results have been reached by combining the large EO functionality of QWs in the InP layer with the photonic integration capability of silicon. The minimal free-carrier freezeout of InP and its alloys is another central enabler to perform efficient modulation in our device with moderate doping, low loss, and wide modulation bandwidth at 4 K. With this active semiconductor platform, we achieve intrinsic Qs as high as 272,000 at 4 K, with much higher Q for larger radius resonators. In this work we demonstrate a single modulator device, but integrating an array of modulators with wavelength multiplexers can increase the aggregate interconnect data bandwidth. Our CRISP modulator can be integrated with CMOS and SC electronics, and it can strongly benefit emerging cryogenic classical and quantum information systems by paving the way for energy-efficient and high-bandwidth optical data-links between 4 K and room temperature, as well as for quantum sensing applications at cryogenic temperatures.

**Funding**

This material is based upon work supported by the Army Research Office (ARO) via IARPA under contract No. W911NF-19-C-0060. Any opinions, findings, and conclusions or recommendations expressed in this material are those of the authors and do not necessarily reflect the views of the Army Research Office.

**Acknowledgments**

We greatly acknowledge the help and support from Mr. Richard Lazarus during the entire execution of the project.

**Supplemental document**

See Supplementary material for supporting content.

**Disclosures**

The authors declare no conflicts of interest.

# Ultralow voltage, High-speed, and Energy-efficient Cryogenic Electro-Optic Modulator

Paolo Pintus,[1,2,‡,*] Anshuman Singh,[3,‡] Weiqiang Xie,[1,‡] Leonardo Ranzani,[3] Martin V. Gustafsson,[3] Minh A. Tran,[1,4] Chao Xiang,[1] Jonathan Peters,[1] John E. Bowers,[1] Moe Soltani[3,*]

Affiliations:
[1]Department of Electrical and Computer Engineering, University of California Santa Barbara, California 93106, USA
[2]Department of Physics, University of Cagliari, Monserrato, 09042, Italy
[3]Raytheon BBN Technologies, 10 Moulton Street, Cambridge, Massachusetts 02138, USA
[4]Currently with Nexus Photonics, Goleta, California 93117, USA

[*]Corresponding authors: ppintus@ece.ucsb.edu, mo.soltani@raytheon.com

[‡]These authors contributed equally to the work

**S1. CRISP resonator Design**

**S2. QW material stack design**

**S3. CRISP modulator fabrication**

**S4. Modulator low frequency and high frequency characterization**

**S5. Characterization of the contact resistance of the modulator**

**S6. Measurement of the resistance (R), capacitance (C) and the electric bandwidth of the modulator**

**S7. I-V curve measurement of the diode of the modulator**

**S8. Demonstration of resonator modulators with intrinsic Q~150,000**

**S1. CRISP resonator design**

The InP-on-Si resonator with all the detailed stack layers were considered for the design and simulation. For all deigns, we considered the transverse-electric (TE) polarization. Figure S1a and S1b shows a typical simulated cross section mode profile for the resonator and the waveguide mode. For the optimal design we considered the followings:

- <u>The minimum width of the InP ring for versatile fabrication</u>: To increase yield of the fabrication of metal electrode on the InP ring we considered a width of 3 micron, though with a more advanced lithography one can consider a InP ring width as small as 2 microns.

- <u>Optimal Si layer dimensions</u>: for maximum index matching and modal overlap between Si and the InP ring: Our simulation showed a Si rib with a total height of 500 nm and partial etch depth of 230 nm provides strong modal overlap.

- <u>InP and Si ring radius</u>: The simulations showed optical mode confinement for ring radii as small as 30 microns. However, for the experiment we fabricated devices with radius ~ 40 micron. A Si ring width of 1000-1200nm guaranteed strong optical confinement in the resonator with negligible radiation leakage.

- <u>Waveguide-resonator coupling</u>: To provide waveguide-resonator coupling with strong resonance extinction, we optimized the waveguide width to satisfy phase matching to the resonator mode. This optimization took the waveguide-resonator spacing (~250nm) into account when optimizing the waveguide width. To further enhance the waveguide-resonator coupling, we used a pulley coupling scheme where the coupling waveguide evanescently raps around the resonator at a pulley angle of ~150°. The Si waveguide layer as shown in Fig. S1b as well as in Fig. 1a (in the main text) is under the InP slab and the bonding layer, and so these layers were considered on top of the waveguide for the simulation.

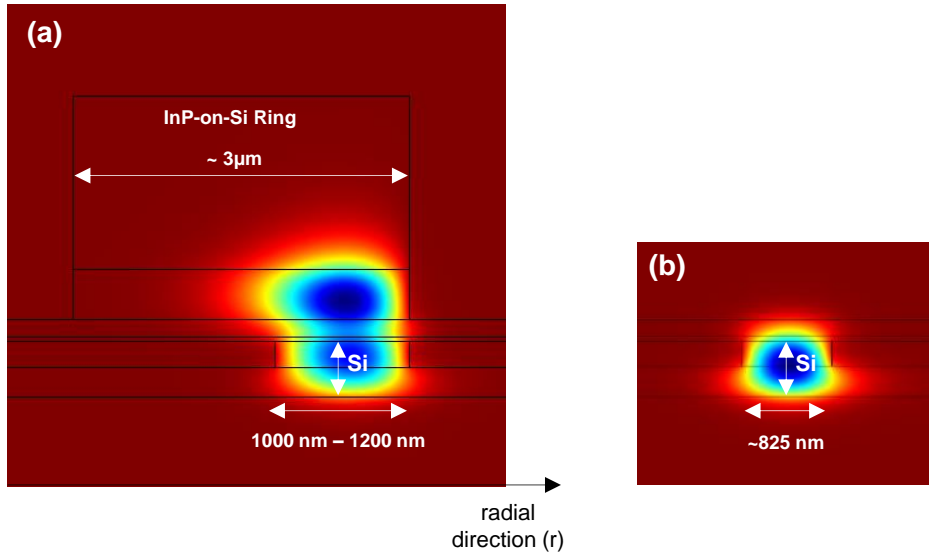

**Fig. S1:** (a) Designed and simulated cross section mode profile of InP-on-Si ring resonator with an external radius of 42 micron. (b) Cross section mode profile of a Si waveguide under the InP slab layer.

**S2. QW material stack design**

A central aspect of our work is in designing the III-V alloy to perform the largest electro-optic (EO) phase shift with minimal optical absorption at 4 K. In a III-V multi QWs, the dominant electro-optic effects that contribute to change of the refractive index are either due to free carriers or due to changing the energy bands when applying an external electric field as described below:

1- The EO effects caused by free carrier are the band-filling (Burstein-Moss) effect, the plasma dispersion effect, and the band-shrinkage effect[1]. Among them, the band-filling is the strongest, while the band shrinkage is the weakest and becomes considerable only at much higher doping concentrations than the one in our case[2].

2- The EO effects due to change of energy band is Quantum Confine Stark effect[3] (QCSE) when applying an electric field.

Note that the III-V alloy also has the Pockels EO effect[4]. However, the contribution of the Pockels effect is small, and in particular in a ring configuration becomes negligible, since all phase shift acquired for propagation around one-half of the ring is reversed during propagation around the other half

A challenge is in designing the III-V alloy material composition for 4 K operation knowing that the low temperature optoelectronic properties of III-V alloys are neither fully characterized nor theoretically modeled in the literature. A significant problem in the theoretical modeling is the sharp transition in the Fermi-Dirac distribution function making numerical calculations prone to instabilities. To overcome these issues, we design the III-V alloy material composition for 4 K operation based on experimental data on modulator devices that were optimized to operate at room temperature[5] and employ the empirical expressions exists for the variation of bandgap energy and lattice constants versus temperatures to optimize the material for 4K operation.

We design the alloy composition of the epitaxial stacks of QW/QB to maximize both the QCSE and the free carrier effects. Figure S2(a)-(c) show the energy band diagram and the energy levels of QW/QB as well as the electron and hole state. More consideration on the energy band gaps of the heterogeneous silicon/III-V phase modulator can be found in H.-W. Chen[6]. To achieve large QCSE, electrons and holes must be confined in the QW layer by optimizing the band offset in the conduction ($\Delta E_c$) and valence ($\Delta E_v$) bands. The band offsets are temperature independent[7,8] and changes solely with the alloy composition. Their values are chosen high enough that both electrons and holes are confined when in their ground states with energy $E_{e,0}$ and $E_{hh,0}$, respectively.

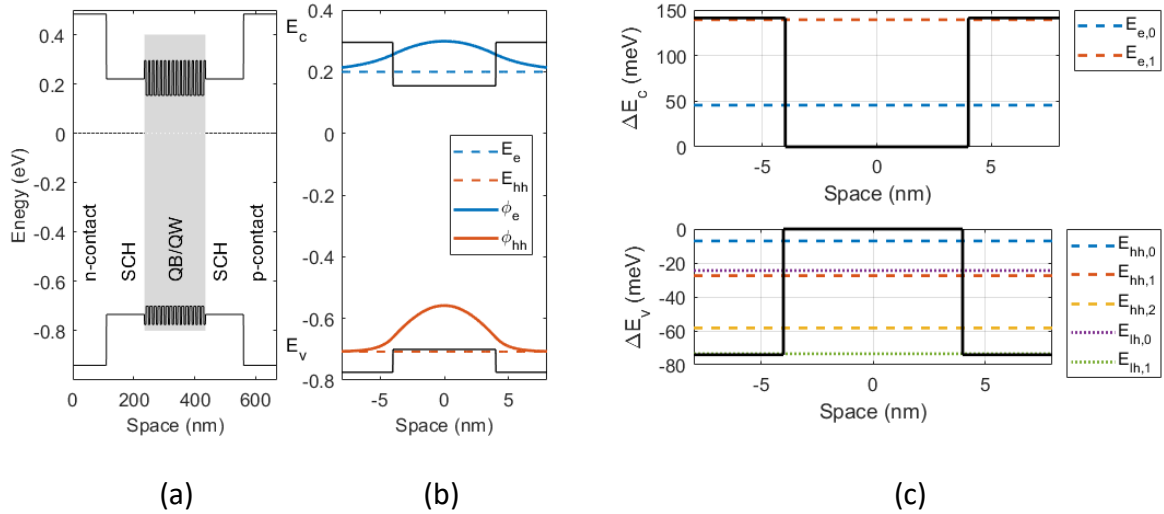

**Fig. S2**: (a) Energy band diagram of the III-V alloy stack, which includes the n- and p-contacts (i.e., InP), the SCH and the multi QW layer (i.e., InAlGaAs). (b) Energy level and wavefunction of the electron and hole in the fundamental energy state. (c) Energy levels of the electron in the conduction band (top), the heavy-hole (dashed) and light-hole (dotted) in the valence band.

When an external voltage is applied, it is desirable to deplete the QW/QB area from electrons and holes, maximizing the refractive index variation due to free carriers. If $\Delta E_c$ and $\Delta E_v$ are too large, the excited states exist and the carriers will stay in the QW rather than being depleted. Since free carrier energy levels depend on the band offset and the QW, we fix $\Delta E_c$ and the well thickness such that only the ground state is occupied by electrons. As a result, $\Delta E_c$ equals to the energy of first excited state ($E_{e,1}$) and the QW thickness is set to 8 nm. On the other hand, the barrier thickness is set equal to 5 nm to make QWs independent.

The value of $\Delta E_v$ and energy level of holes are defined from the previous constraint. Minimizing the optical absorption is another important design goal in the modulator. Significant absorption occurs when the photon energy of the injected light approaches the exciton binding energy of the QW, $\Delta E_{PL}=E_{e,0}-E_{hh,0}$. To minimize the optical absorption , the exciton binding energy (i.e., photoluminescence) is set ~100 meV larger than the energy of operating wavelength[9] ~800 meV (λ~1550 nm). From those consideration, the optimal band gaps (in wavelength) for the SCH, QW and QB are $\lambda_{g,SCH}$=1.3 μm, $\lambda_{g,QW}$=1.16 μm, $\lambda_{g,QB}$=1.46 μm, respectively.

The material composition of QW, QB and SCH are determined from Fig. S3, which shows the band gap and the strain at 4 K of InAlGaAs as a function of the indium (x) and gallium (y) concentrations in the alloy. The domain of the feasible material composition is $0 \leq x \leq 1$, $0 \leq y \leq 1$ and $0 \leq 1 - x - y$. The area above $1 - x - y = 0$ is unfeasible since the concentration of aluminum becomes negative (dash dot line in the figure). For low concentrations of In and Ga, the bandgap of InAlGaAs is indirect, which is undesirable (dashed line). Thus, the region of interest falls between the dashed and the dot-dashed black line. The black dotted line indicates material compositions that match the lattice of the underlying InP, which is preferred in order to avoid stress in the material at cryogenic temperatures.

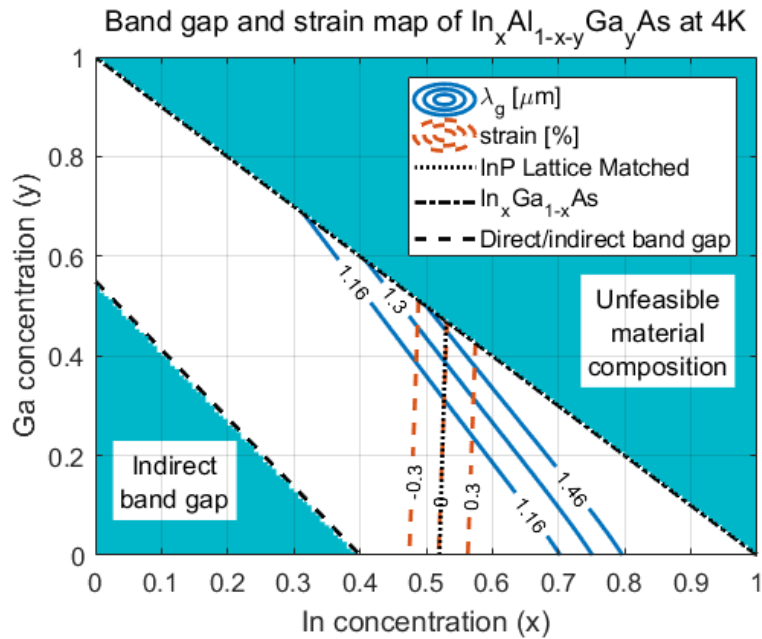

**Fig. S3**: Band gap in wavelength (blue solid line) and lattice strain relative to InP (red dashed line) for InAlGaAs with respect to indium (x) and gallium (y) concentrations. The feasible alloy compositions are those where the cumulative concentration of indium and gallium are below 1 (black dot dashed like). In the same plot, we indicated the transition between direct and indirect band gap (black dashed line), and the alloy composition that matches the lattice of InP (black dotted line).

The contour lines of constant band gap energies (blue solid lines) for a temperature T=4 K are computed interpolating the bang gap energies of InAs, AlAs and GaAs at the same temperature[10,11]. Specifically, the energy gap at 4 K of such binary compounds is calculated using the Varshni's empirical equation

$$E_g(T) = E_g(0) - \frac{\alpha T^2}{T + \beta} \qquad (S1)$$

$$\lambda_g(T) = \frac{hc}{E_g(T)} \qquad (S2)$$

where, $E_g(T)$ is the temperature dependent band gap, $\lambda_g(T)$ is the corresponding wavelength, $E_g(0)$, α and β material constants, h the Plank constant, and c the speed of light in the vacuum. For InP, InAs, AlAs and GaAs the constants, $E_g(0)$, α and β by are reported in Table S1.

Table S1: Fitting parameters of the Varshni's empirical equation for the alloy used in CRISP modulator[7]. From Eq.(S1), the energy band gap at 4K can be easily computed.

| Material | $E_g(0)$ | α | β |
|---|---|---|---|
| InP | 1.4236 | 0.363·10⁻³ | 162 |
| InAs | 0.417 | 0.276·10⁻³ | 93 |
| AlAs | 3.099 | 0.885·10⁻³ | 530 |
| GaAs | 1.519 | 0.541·10⁻³ | 204 |

The lattice mismatch between InAlGaAs and InP native substrate is shown in the same figure (red dotted line in Fig. S3). We seek a material composition with minimal strain at room temperature, which can be exacerbated at low temperature and diminish the performance of the material. As a result, we obtain the following material composition for the SCH (Ga= 0.386, In=0.529), the QW (Ga= 0.449, In=0.530) and the QB (Ga= 0.312, In=0.527), which are determined by the intersection of the three blue lines and the black dot curve in Fig. S3

The epitaxial layers are grown one after the other on the same substrate using metal organic chemical vapor deposition (MOCVD) technique. If the concentration of all the elements (i.e., In, Al, and Ga) are varied during the growth process, it is difficult to control the growth conditions, and the final composition may deviate from the original design. Since the concentration of indium does not change significantly, we set the in concentration equal to the average value 0.5296, while the concentrations of Ga and Al are adjusted to achieve the desired energy gap. The final stack and alloy composition are reported in Fig S4. Beyond the SCH, QW and QB, the composition of the other layers is similar to the one used for other heterogeneous III-V/silicon photonics[6,12,13].

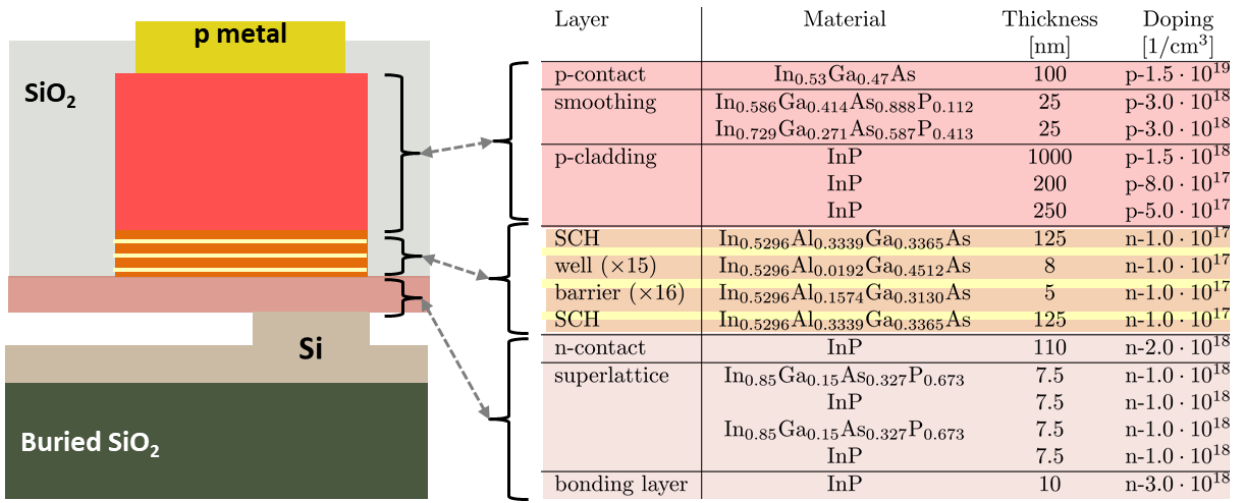

**Fig S4**: III/V epitaxial layers used for the cryogenic QW based modulator

### S3. CRISP modulator fabrication

The major fabrication steps of heterogeneous InP-on-Si modulator is depicted in Figure S5. It consists of silicon patterning and etching, III-V bonding and processing, followed by metal deposition and rapid thermal annealing for the contacts. All lithographic steps are performed using a 248 nm deep-ultra violet (DUV) stepper. The entire fabrication takes place at a wafer scale (4" SOI wafer), which demonstrates the potential scalability of the process.

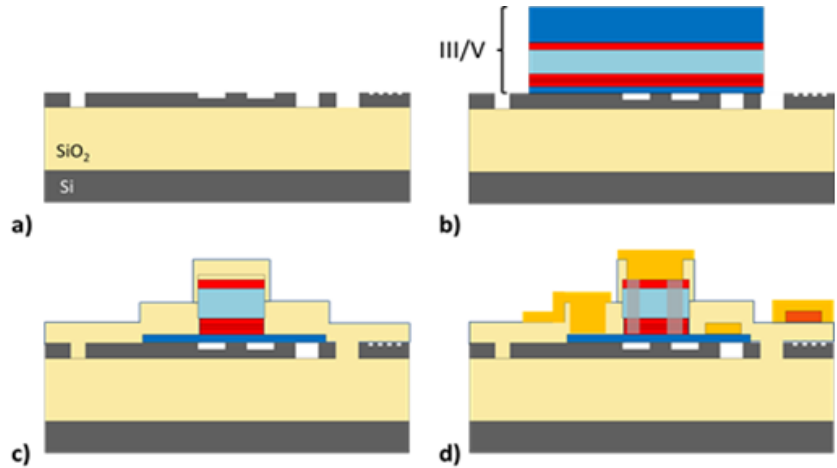

**Fig. S5**: Major steps of the InP-on-Si modulator fabrication (a) Silicon fabrication, (b) InP-on-Si bonding, (c) InP patterning and cladding, (d) Metallization and defining the contact electrodes.

The first part of the process consists on the patterning of the silicon layer for the fabrication of the silicon waveguide, ring resonator and edge couplers. Due to the multiple silicon etching steps with different etch depths, a SiO2 hard-mask is used to define the waveguides. The etching process is then carried out with the shallowest etch (240 nm) for defining the ring and waveguide. A second deep etch step is used to remove the left silicon to define the edge coupler (500 nm). Since the hard-mask is used to define the waveguides during the entire process, it ensures that the transitions between waveguides with different etch depths are self-aligned and not subject to lithographic misalignment in the stepper.

The III-V die is bonded using an $O_2$ plasma activated direct bonding procedure, and the bulk of the InP substrate is removed using a combination of lapping and wet etching, leaving the III/V epitaxial layers used for the cryogenic modulator The modulator mesas are etched using a series of both $CH_4/H_2/Ar$ dry etches and wet etches containing HCl or $H_3PO_4$. Figure S6 shows the SEM images of the InP-on-Si resonator device before adding the oxide cladding and the metallization process. After the III-V processing is complete, SiO$_2$ is sputtered, and vias are opened to expose the n and p-contacts.

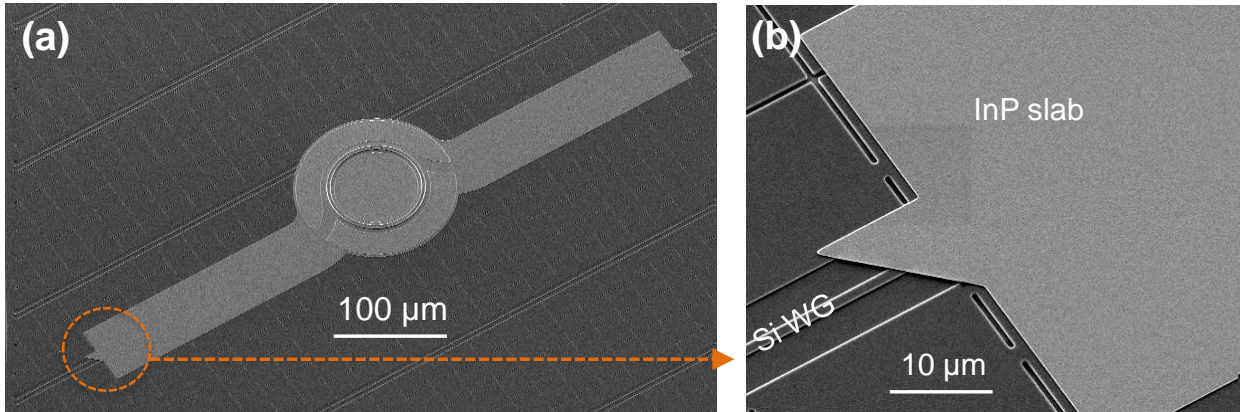

**Fig. S6**: (a) Scanning electron microscope image of the InP-on-Si ring before the stapes of metallization. (b) A zoomed view of the transition region from the Si waveguide to the InP-on-Si region. The transition region has been designed in that shape to minimize back reflection and improve the mode conversion from the silicon waveguide to the heterogeneous cross-section[12]

The last part of the fabrication is deposition of the various metals for n-(Pd/Ge/Pd/Au/Ti) and p-contacts (Pd/Ti/Pd/Au), and probe pads (Au/Ti). More details on the fabrication of heterogeneous silicon photonic devices are discussed in M. Davenport[12].

**S4. Modulator characterization**

The CRISP modulators are characterized in a cryogenic probe station (Montana Cryostation s200) with piezoelectric actuator stages for controlling the optical fibers that couple light in and out the chip, as well as the electrical probe that is used for both DC and RF measurements. The temperature is monitored using a sensor in the proximity of the modulator chip.

The layout of the experimental setup is schematically shown in Fig. S7. The laser light from a tunable laser source is routed through a polarization adjuster to a lensed fiber inside the probe station, which couples it to a silicon waveguide on the CRISP chip. Light is collected from the other facet of the chip using a second lensed fiber and brought outside the cryostat for amplification, photodetection, and further processing in the electrical domain. By sweeping the laser wavelength and monitoring the optical power at the detector, we find the resonance spectrum of the modulator. We repeat this experiment for different bias voltages applied to the modulator to find the resonance shift sensitivity.

For high frequency electro-optic characterization ($S_{21}$), the laser light is held fixed at the full-width at half-maximum (FWHM) of one of the resonances, when a bias DC voltage is applied through a bias-T. The RF signal is applied to the modulator from a vector network analyzer (VNA), and the modulated light is then amplified at the output facet of the CRISP chip using an optical pre-amplifier (Nuphoton) and send to a high-speed photodetector (Finisar). A tunable bandpass filter is used after the optical amplifier to suppress the amplifier background noise. The output of the photodetector is then sent to the input port of the VNA.

For the eye diagram measurement, the random bit sequences are sent to the modulator from a pseudo random bit sequencer ($2^7$-1) at different data rates. The modulated light is detected by a high-speed photodetector, which converts the optical signal into an electrical one. Finally, the electrical signal is sent to a sampling oscilloscope for the eye diagram patterns.

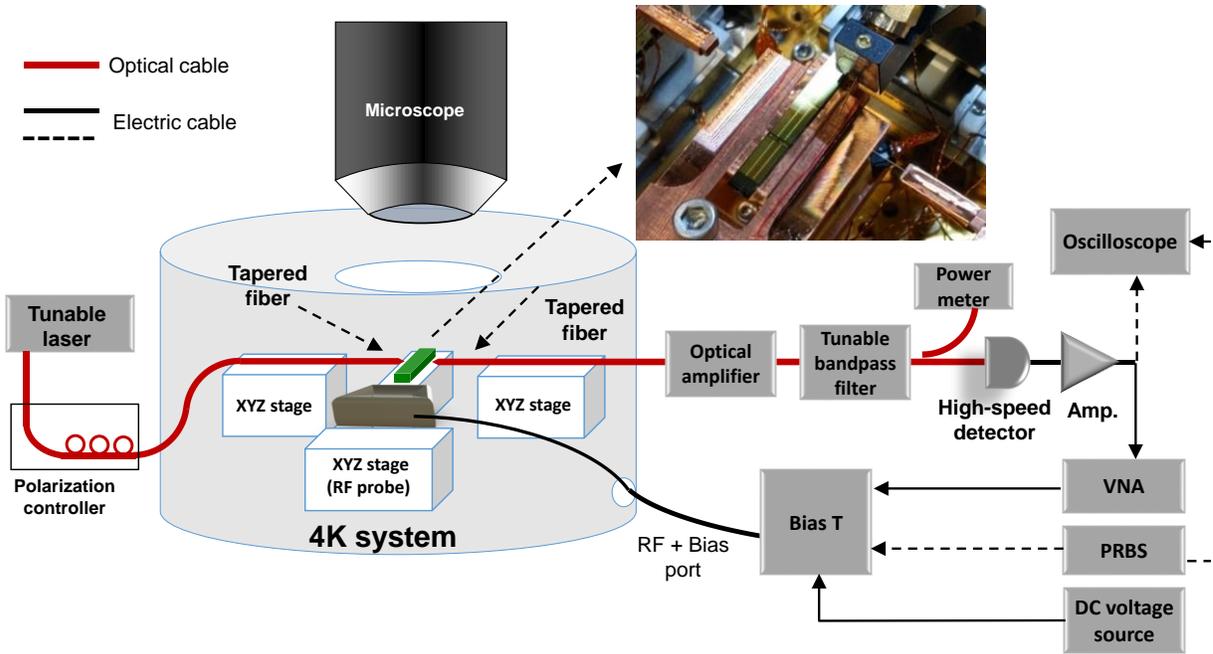

**Fig. S7**: Measurement set-up for the bandwidth and the eye diagram measurements.

**S5. Characterization of the contact resistance of the modulator**

The semiconductor sheet resistance and the semiconductor-metal contact resistance are characterized at 300 K and at 4 K using the transfer length method (TLM). The TLM test pattern consists of an array of identical contacts with various spacings[14]. The measured resistances between neighboring contacts are shown in Fig. S6(a) and Fig. S6(b), for n- and p-type contacts, respectively. The sheet resistance, $R_{sh}$, and the contact resistance, $R_c$, are the slope and the y-intercept divided by 2 of the line that fit the experimental data. For the sake of completeness, we computed also the transfer length, $L_T$, and the contact resistivity, $\rho_c = L_T^2 \cdot R_{sh}$, of the semiconductor-metal interface. The experimental results are reported in Table S2 where no significant change can be observed between room temperature and T= 4 K, which confirms our assumption that III-V are not significantly affected by carrier freezing.

Table S2: Measured contact resistance ($R_c$) and the resistivity ($\rho_c$) of the n- and p- regions of the CRISP modulator. Transfer length ($L_T$) and contact resistivity ($\rho_c$) are also computed.

| Contact | $R_{sh}$ [Ω/□] | $R_c$ [Ω] | $L_T$ [μm] | $\rho_c$ [Ω·cm$^2$] | T[K] |
|---|---|---|---|---|---|
| n | 172.04 | 6.28 | 7.30 | $9.17 \cdot 10^{-5}$ | 300K |
| n | 177.94 | 12.50 | 14.06 | $3.51 \cdot 10^{-4}$ | 4K |
| p | 348.19 | 1.95 | 1.12 | $4.35 \cdot 10^{-6}$ | 300K |
| p | 298.36 | 2.09 | 1.40 | $5.84 \cdot 10^{-6}$ | 4K |

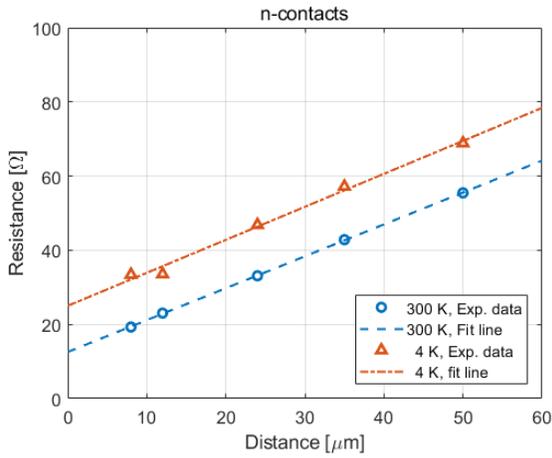

(a)

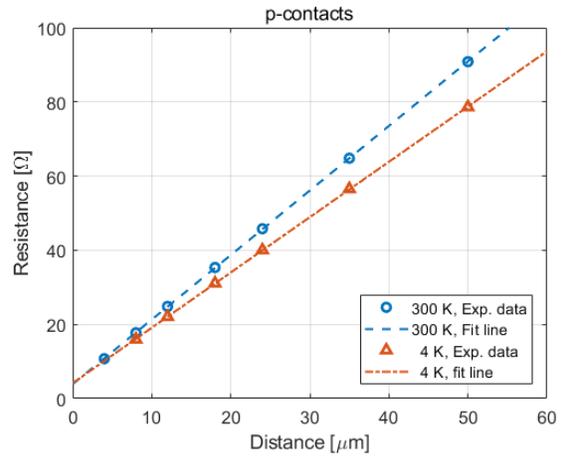

(b)

Fig. S8: TLM characterization of the n- and p- contacts at room temperature and at 4 K

## S6. Measurement of the resistance (R), capacitance (C), and the electric bandwidth of the modulator

To evaluate the energy-per-bit of the CRISP modulator, the resistance, R, and the capacitance, C, of the device are measured at 4 K. The value of R and C are derived from the measurements of the electrical reflection coefficient, $S_{11}$, knowing that

$$S_{11}(\omega) = \frac{\left(R + \frac{1}{j\omega C} - Z_0\right)}{\left(R + \frac{1}{j\omega C} + Z_0\right)} \tag{S3}$$

where $Z_0$ = 50 Ω is the characteristic impedance of the transmission line, and ω is the angular frequency. Coefficient $S_{11}$ is measured with a vector network analyzer at 4 K after the setup was calibrated (including the probe tips). The experimental results are reported in the Table S2.

**Table S2**: Measured capacitance and the resistance of the CRISP modulator with respect to the bias voltage. The measurements are performed at 4 K.

| Bias [V] | 0.5 | 0 | -1.0 | -2.0 | -3.0 | -4.0 | -5.0 | -6.0 |
|---|---|---|---|---|---|---|---|---|
| C [fF] | 418 | 406 | 406 | 406 | 407 | 411 | 414 | 417 |
| R [Ω] | 16 | 17.4 | 17.1 | 17.1 | 17 | 16.9 | 16.9 | 16.9 |

In this device, we estimate an electrical bandwidth $BW_e = 1/2\pi (R+Z_0) C = 6$ GHz, where $Z_0$ = 50 Ω is the characteristic impedance of the transmission line. A bandwidth larger than 25 GHz can be achieved by lowering the capacitance. This can be done by reducing the size of the electrical contact pads of the modulator and introducing an intrinsic region between the p and n doped area (p-i-n configuration[15]).

The dynamic electric energy per bit consumption of the device can be computed from[16–18]

$$E_b = \frac{1}{4} CV^2 \tag{S4}$$

with $V_{pp}$ = ~10 mV, we get ~10.4 attojoule/bit energy consumption.

**S7. I-V curve measurement of the diode of the modulator**

Figure S9 shows the measured I-V curve of the modulator at 4 K. For this measurement, a swept scan voltage from a Keysight precision source measurement unit (2911B) is applied to the modulator electrode pads, and the current is read by the same measurement unit. As seen from

Fig.S9a, a typical diode I-V curve is observed for the modulator. Figure S9b shows a zoomed view at the reverse bias region wherein a leakage current in the order of 0.2-0.4 nA is observed. The static electric energy consumption of the modulator is calculated as $|V_{bias}I_{bias}|/r_b$, where $V_{bias}$, $I_{bias}$, and $r_b$ are the DC bias voltage, the bias (leakage) current, and the bit rate respectively.

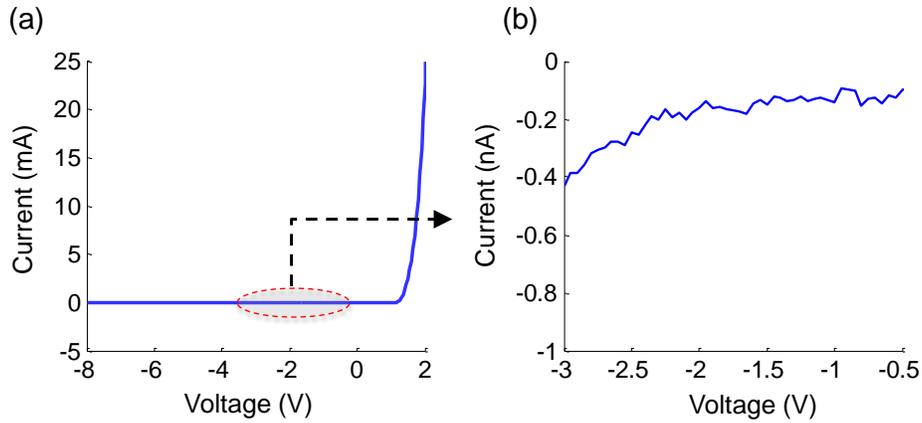

**Fig. S9:** Measured I-V curve of the diode of the resonator modulator at 4 K temperature

### S8. Demonstration of resonator modulators with intrinsic Q~150,000

Our CRISP platform enables the demonstration of high Q resonator modulators at cryogenic temperatures. Fig. S10 shows the measured results of one of such resonator modulators. Fig. S10(a) shows the resonance spectrum for applied bias voltages of 0 V and -0.5 V and strong extinction close to critical coupling. For the -0.5 V bias, the resonance extinction is 18 dB, and the linewidth at 3 dB below the maximum transmission is ~10.3 pm (see the black dashed line on red resonance spectrum) corresponding to a loaded quality factor $Q_L$ = 153,000. Knowing the resonance extinction and the loaded quality factor we can calculate the intrinsic quality factor ($Q_0$) using this expression[19]:

$$Q_0 = \frac{2Q_L}{\pm\sqrt{T}+1} \tag{S5}$$

In the above expression T is the extinction at the resonance and its value is 0.016 (~18 dB). Knowing that the resonator is slightly under-coupled compared to the critical coupling regime, we take the plus sign in the expression in Eq. (S5). As a result, we find an intrinsic Q of ~272,000 This number is a record for semiconductor resonator modulators. In principle, we can achieve higher intrinsic Qs by going to larger ring radii without needing to lower the doping level. Depending on the application and the required modulation bandwidth, we can also optimize the doping level to achieve higher Qs approaching the million range.

Figure S10(b) shows the modulator frequency response with a 3 dB bandwidth of ~1.2 GHz indicating that this device can be used for high-speed nodulation.

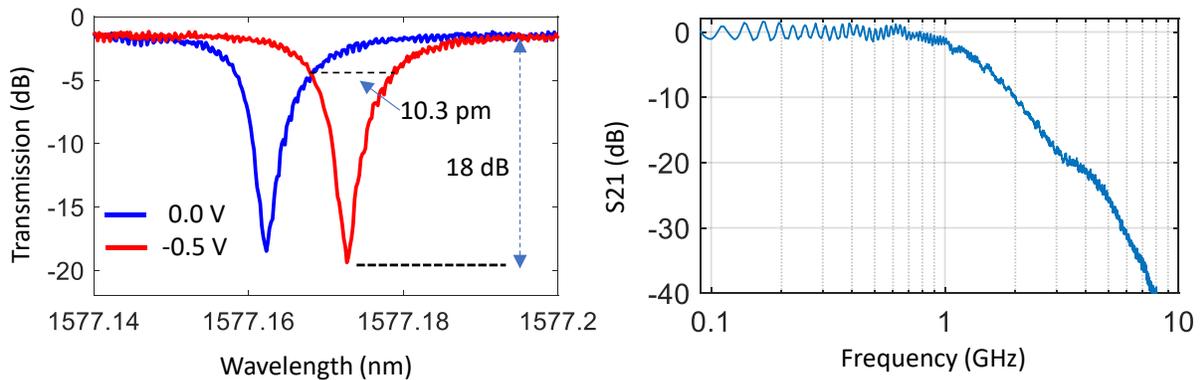

**Fig. S10**: **Measured high Q resonator modulators at 4 K temperature**. (a) Resonator spectrum close to critical coupling (18 dB extinction), with a loaded linewidth of ~10.3 pm, corresponding to a loaded Q of 153,000 and an intrinsic Q of 72,000 which is a record for a semiconductor resonator modulator device. (b) Measured modulator frequency response with a 3 dB bandwidth of 1.2 GHz.